# Massive Black Holes: Evidence, Demographics and Cosmic Evolution

**Reinhard Genzel**
*Max-Planck Institut für Extraterrestrische Physik (MPE), Garching, FRG*
*& Departments of Physics & Astronomy, University of California, Berkeley, USA*

The article summarizes the observational evidence for the existence of massive black holes, as well as the current knowledge about their abundance, their mass and spin distributions, and their cosmic evolution within and together with their galactic hosts. We finish with a discussion of how massive black holes may in the future serve as laboratories for testing the theory of gravitation in the extreme curvature regimes near the event horizon.

## 1. Introduction

In 1784 Rev. John Michell was the first to note that a sufficiently compact star may have a surface escape velocity exceeding the speed of light. He argued that an object of the mass of the Sun (or larger) but with a radius of 3 km (instead of the Sun's radius of 700,000 km) would thus be invisible. A proper mathematical treatment of this problem then had to await Albert Einstein's General Relativity ("GR", 1916). Karl Schwarzschild's (1916) solution of the vacuum field equations in spherical symmetry demonstrated the existence of a characteristic event horizon, the Schwarzschild radius $R_s=2GM/c^2$, within which no communication is possible with external observers. Roy Kerr (1963) generalized this solution to spinning black holes. The mathematical concept of a black hole was established (although the term itself was coined only later by John Wheeler in 1968). In GR, all matter within the event horizon is predicted to be inexorably pulled toward the center where all gravitational energy density (matter) is located in a density singularity. From considerations of the information content of black holes, there is significant tension between the predictions of GR and Quantum theory (e.g. Susskind 1995, Maldacena 1998, Bousso 2002). It is generally thought that a proper quantum theory of gravity will modify the concepts of GR on scales comparable to or smaller than the Planck length, $l_{Pl}$~$1.6x10^{-33}$cm, remove the concept of a central singularity, and potentially challenge the interpretation of the GR event horizon (Almheiri et al. 2013).

But are these objects of GR realized in Nature?

## 2. First Evidence

Astronomical evidence for the existence of black holes started to emerge in the 1960s with the discovery of distant luminous 'quasi-stellar-radio-sources/objects' (QSOs, Schmidt 1963) and variable X-ray emitting binaries in the Milky Way (Giacconi et al. 1962). It became clear from simple energetic arguments that the enormous luminosities and energy densities of QSOs (up to several $10^{14}$ times the luminosity of the Sun, and several $10^{4}$ times the entire energy output of the Milky Way Galaxy), as well as their strong UV-, X-ray and radio emission can most plausibly be explained by accretion of matter onto massive black holes (e.g. Lynden-Bell 1969, Shakura & Sunyaev 1973, Rees 1984, Blandford 1999). Simple theoretical considerations show that between 7% (for a non-rotating Schwarzschild hole) and 40% (for a maximally rotating Kerr hole) of the rest energy of an infalling particle can in principle be converted to radiation outside the event horizon, a factor 10 to 100 more than in stellar fusion from hydrogen to helium. To explain powerful quasars by this mechanism, black hole masses of $10^{8}$ to $10^{9}$ solar masses and accretion flows between 0.1 to 10 solar masses per year are required. QSOs are located (without exception) at the nuclei of large, massive galaxies (e.g. Osmer 2004). QSOs just represent the most extreme and spectacular among the general nuclear activity of most galaxies. This includes variable X- and γ-ray emission and highly collimated, relativistic radio jets, all of which cannot be accounted for by stellar activity.

The 1960s and 1970s brought also the discovery of X-ray stellar binary systems (see Giacconi 2003 for an historic account). For about 20 of these compact and highly variable X-ray sources dynamical mass determinations from Doppler spectroscopy of the visible primary star established that the mass of the X-ray emitting secondary is significantly larger than the maximum stable neutron star mass, ~3 solar masses (McClintock & Remillard 2004, Remillard & McClintock 2006, Özel et al. 2010). The binary X-ray sources thus are excellent candidates for stellar black holes (SBH). They are probably formed when a massive star explodes as a supernova at the end of its fusion lifetime and the compact remnant collapses to a stellar hole.

An unambiguous proof of the existence of a stellar or massive black hole, as defined by GR, requires the determination of the gravitational potential to the scale of the event horizon. This proof can in principle be obtained from spatially resolved measurements of the motions of test particles (interstellar gas or stars) in close orbit around the black hole. In practice it is not possible (yet) to probe the scale of an event horizon of any black hole candidate (SBH as well as MBH) with spatially resolved dynamical measurements. A more modest goal then is to show that the gravitational potential of a galaxy nucleus is dominated by a compact non-stellar mass and that this central mass concentration cannot be anything but a black hole because all other conceivable configurations are more extended, are not stable, or produce more light (e.g. Maoz 1995, 1998). Even this test cannot be conducted yet in distant QSOs from dynamical measurements. It has become feasible over the last decades in nearby galaxy nuclei, however, including the Center of our Milky Way.

## 3. NGC 4258

Solid evidence for central ‘dark’ (i.e. non-stellar) mass concentrations in about 80 nearby galaxies has emerged over the past two decades (e.g. Magorrian 1998, Kormendy 2004, Gültekin et al. 2009, Kormendy & Ho 2013, McConnell & Ma 2013) from optical/infrared imaging and spectroscopy on the Hubble Space Telescope (HST) and large ground-based telescopes, as well as from Very Long Baseline radio Interferometry (VLBI).

The first truly compelling case that such a dark mass concentration cannot just be a dense nuclear cluster of white dwarfs, neutron stars and perhaps stellar black holes emerged in the mid-1990s from spectacular VLBI observations of the nucleus of NGC 4258, a mildly active galaxy at a distance of 7 Mpc (Miyoshi et al. 1995, Moran 2008, Figure 1). The VLBI observations show that the galaxy nucleus contains a thin, slightly warped disk of $H_2O$ masers (viewed almost edge on) in Keplerian rotation around an unresolved mass of 40 million solar masses (Figure 1). The inferred density of this mass exceeds a few $10^9$ solar masses pc$^{-3}$ and thus cannot be a long-lived cluster of ‘dark’ astrophysical objects of the type mentioned above (Maoz 1995). As we will discuss below, a still more compelling case can be made in the case of the Galactic Center.

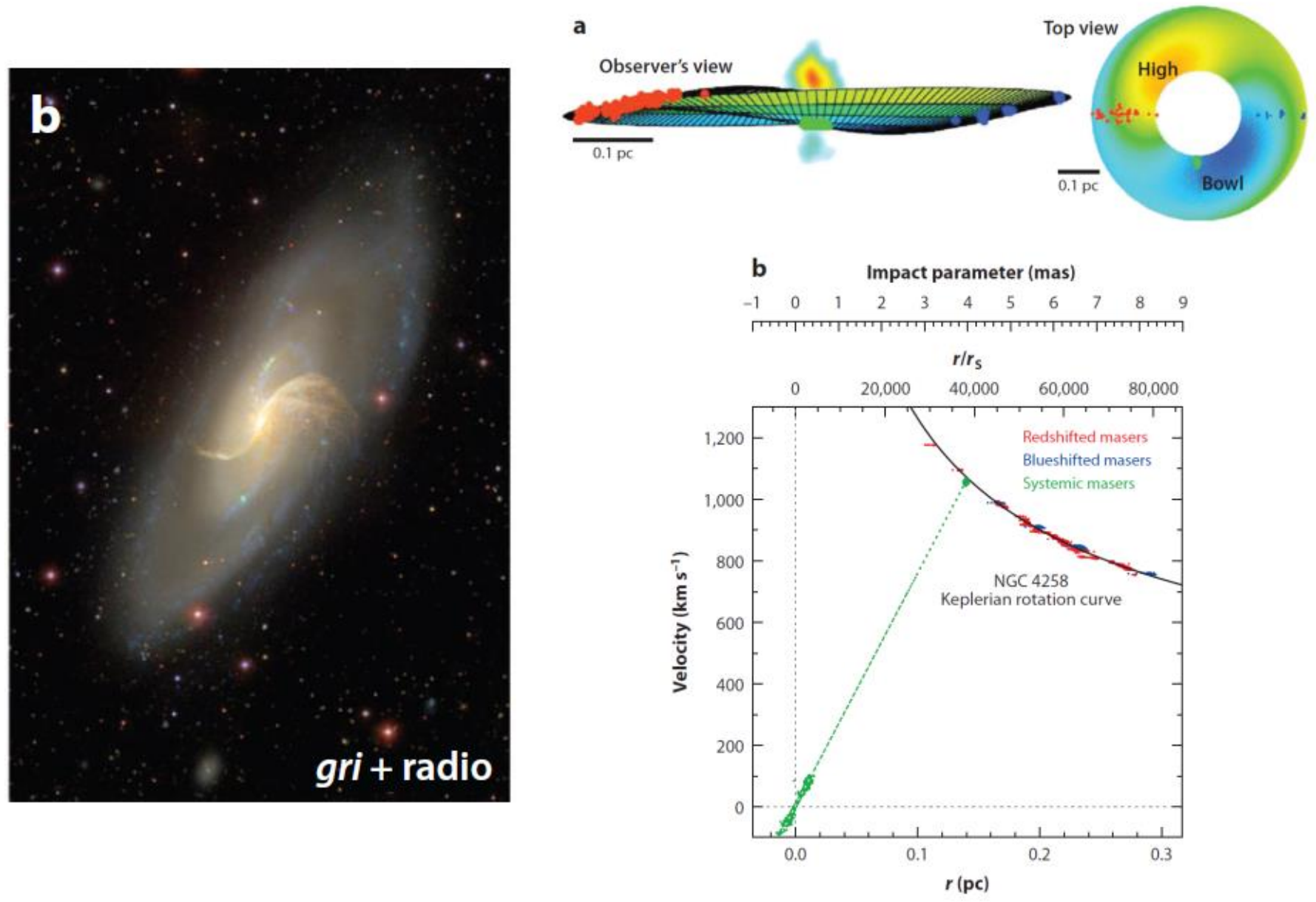

Figure 1. Left: Optical and radio image of the active galaxy NGC4258. This disk galaxy exhibits a spectacular curved twin radio and X-ray jet, visible in orange in this picture. Right: (top) Schematic edge-on (left) and face-on (right) views of the almost-edge-on, warped maser disk of NGC 4258 (from Moran 2008) with warp parameters from Herrnstein et al. (2005) and including the inner contours of the radio jet. The relative positions of the receding, near-systemic, and approaching $H_2O$ masers are indicated by red, green, and blue spots, respectively. Differences in line-of-sight projection corrections to the slightly tilted maser velocities account for the departures in the high-velocity

masers from exact Keplerian rotation. The near-systemic masers are seen tangent to the bottom of the maser disk bowl along the line of sight. They drift from right to left in ~ 12 years across the green areas where amplification of the background radio continuum is sufficient for detection. (b) NGC 4258 rotation velocity versus radius in units of parsec (bottom axis), Schwarzschild radii (top axis), and milliarcsec (extra axis). The black curve is a Keplerian fit to 4255 velocities of red- and blue-shifted masers (red and blue dots). The small green points and line show 10036 velocities of near-systemic masers and a linear fit to them. The green filled circle is the corresponding mean velocity point. The maser data are taken from Argon et al. (2007) (adapted from Kormendy & Ho 2013).

## 4. The Galactic Center Black Hole

The central light years of our Galaxy contain a dense and luminous star cluster, as well as several components of neutral, ionized and extremely hot gas (Genzel, Hollenbach & Townes 1994, Genzel, Eisenhauer & Gillessen 2010). The central dark mass concentration discussed above is associated with the compact radio source SgrA*, which has a size of about 10 light minutes and is located at the center of the nuclear star cluster. SgrA* thus may be a MBH analogous to QSOs, albeit with orders of magnitude lower mass and luminosity. Because of its proximity - the distance to the Galactic Center is about 8.3 kilo-parsecs (kpc), about $10^5$ time closer than the nearest QSOs - high resolution observations of the Milky Way nucleus offer the unique opportunity of carrying out a stringent test of the MBH-paradigm and of studying stars and gas in the immediate vicinity of a MBH, at a level of detail that will not be accessible in any other galactic nucleus for the foreseeable future. Since the Center of the Milky Way is highly obscured by interstellar dust particles in the plane of the Galactic disk, observations in the visible part of the electromagnetic spectrum are not possible. The veil of dust, however, becomes transparent at longer wavelengths (the infrared, microwave and radio bands), as well as at shorter wavelengths (hard X-ray and $\gamma$-ray bands), where observations of the Galactic Center thus become feasible.

The key obviously lies in very high angular resolution observations. The Schwarzschild radius of a 4 million solar mass black hole at the Galactic Center subtends a mere $10^{-5}$ arc-seconds[1]. For high resolution imaging from the ground an important technical hurdle is the correction of the distortions of an incoming electromagnetic wave by the refractive Earth atmosphere. For some time radio astronomers have been able to achieve sub-milli-arcsecond resolution VLBI at millimeter wavelengths, with the help of phase-referencing to nearby compact radio sources. In the optical/near-infrared waveband the atmosphere distorts the incoming electromagnetic waves on time scales of milliseconds and smears out long-exposure images to a diameter of more than an order of magnitude greater than the diffraction limited resolution of large ground-based telescopes (Figure 2). From the early 1990s onward initially 'speckle imaging' (recording short exposure images, which are subsequently processed and co-added to retrieve the diffraction limited resolution) and then later ‘adaptive optics’ (AO: correcting the wave distortions on-line) became available, which have since allowed increasingly precise high resolution near-infrared observations with the currently largest (10 m diameter) ground-based telescopes of the Galactic Center (and nearby galaxy nuclei).

[1] 10 μarc-seconds correspond to about 2cm at the distance of the Moon

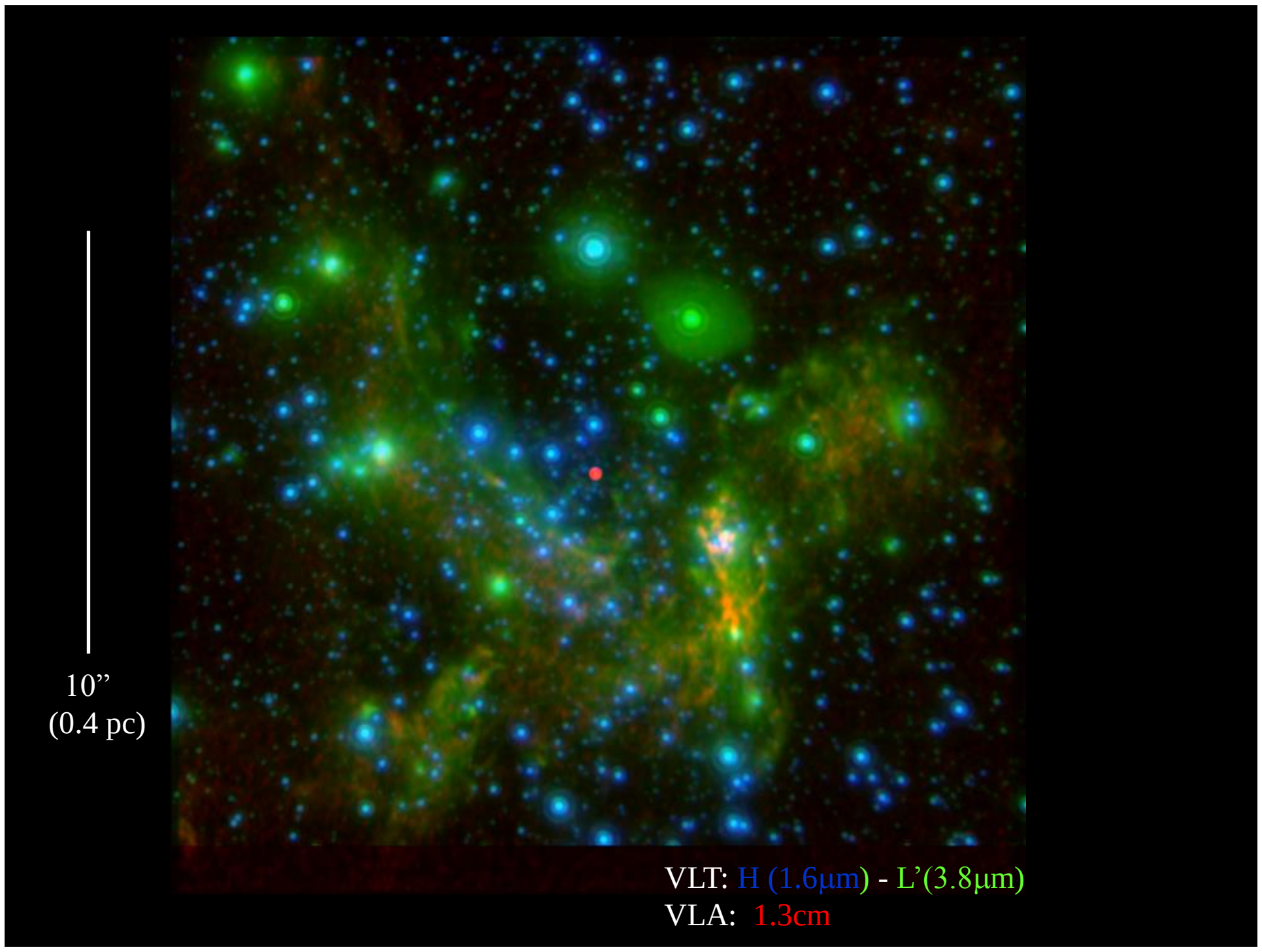

Figure 2. Near-infrared/radio, color-composite image of the central light years of Galactic Center. The blue and green colors represent the 1.6 and 3.8µm broad band near-infrared emission, at the diffraction limit (~0.05”) of the 8m Very Large Telescope (VLT) of the European Southern Observatory (ESO), and taken with the ‘NACO’ AO-camera and an infrared wavefront sensor (adapted from Genzel et al. 2003). Similar work has been carried out at the 10 m Keck telescope (Ghez et al. 2003, 2005). The red color image is the 1.3cm radio continuum emission taken with the Very Large Array (VLA) of the US National Radio Astronomy Observatory (NRAO). The compact red dot in the center of the image is the compact, non-thermal radio source SgrA*. Many of the bright blue stars are young, massive O/B- and Wolf-Rayet stars that have formed recently. Other bright stars are old, giants and asymptotic giant branch stars in the old nuclear star cluster. The extended streamers/wisps of 3.8µm emission and radio emission are dusty filaments of ionized gas orbiting in the central light years (adapted from Genzel, Eisenhauer & Gillessen 2010).

Early evidence for the presence of a non-stellar mass concentration of 2-4 million times the mass of the Sun ($M_\odot$) came from mid-infrared imaging spectroscopy of the 12.8µm [NeII] line, which traces emission from ionized gas clouds in the central parsec region (Wollman et al. 1977, Lacy et al. 1980, Serabyn & Lacy 1985). However, many considered this dynamical evidence not compelling because of the possibility of the ionized gas being affected by non-gravitational forces (shocks, winds, magnetic fields). A far better probe of the gravitational field are stellar motions, which started to become available from Doppler spectroscopy in the late 1980s. They confirmed the gas motions (Rieke & Rieke 1988, McGinn et al. 1989, Sellgren et al. 1990, Krabbe et al. 1995, Haller et al. 1996, Genzel et al. 1996). The ultimate breakthrough came from the combination of AO techniques with advanced imaging and spectroscopic instruments (e.g. 'integral field' imaging spectroscopy, Eisenhauer et al. 2005) that allowed diffraction limited near-infrared spectroscopy and imaging astrometry with a precision initially at the few milli-arcsecond scale, and improving to a few hundred micro-arcseconds in the next decade (c.f. Ghez et al. 2008, Gillessen et al. 2009). With diffraction limited imagery starting in 1992 on the 3.5m New Technology Telescope (NTT) of the European Southern Observatory (ESO) in La Silla/Chile, and continuing since 2002 on ESO's Very Large Telescope (VLT) on Paranal, a group at MPE was able to determine proper motions of stars as close as ~0.1" from SgrA* (Eckart & Genzel 1996, 1997). In 1995 a group at the University of California, Los Angeles started a similar program with the 10m diameter Keck telescope in Hawaii (Ghez et al. 1998). Both groups independently found that the stellar velocities follow a 'Kepler' law ($v \sim R^{-1/2}$) as a function of distance from SgrA* and reach $\geq 10^3$ km/s within the central light month.

Only a few years later both groups achieved the next and crucial steps. Ghez et al. (2000) detected accelerations for three of the 'S'-stars, Schödel et al. (2002) and Ghez et al. (2003) showed that the star S2/S02 is in a highly elliptical orbit around the position of the radio source SgrA*, and Schödel et al. (2003) and Ghez et al. (2005) determined the orbits of 6 additional stars. In addition to the proper motion/astrometric studies, they obtained diffraction limited Doppler spectroscopy of the same stars (Ghez et al.2003, Eisenhauer et al. 2003, 2005), allowing precision measurement of the three dimensional structure of the orbits, as well as the distance to the Galactic Center. Figure 3 shows the data and best fitting Kepler orbit for S2/S02, the most spectacular of these stars with a 16 year orbital period (Ghez et al. 2008, Gillessen et al. 2009, 2009a). At the time of writing, the two groups have determined individual orbits for more than 40 stars in the central light month. These orbits show that the gravitational potential indeed is that of a point mass centered on SgrA*. These stars orbit the position of the radio source SgrA*like planets around the Sun. The point mass must be concentrated well within the peri-approaches of the innermost stars, ~10-17 light hours, or 70 times the Earth orbit radius and about 1000 times the event horizon of a 4 million solar mass black hole. There is presently no indication for an extended mass greater than about 2 % of the point mass.

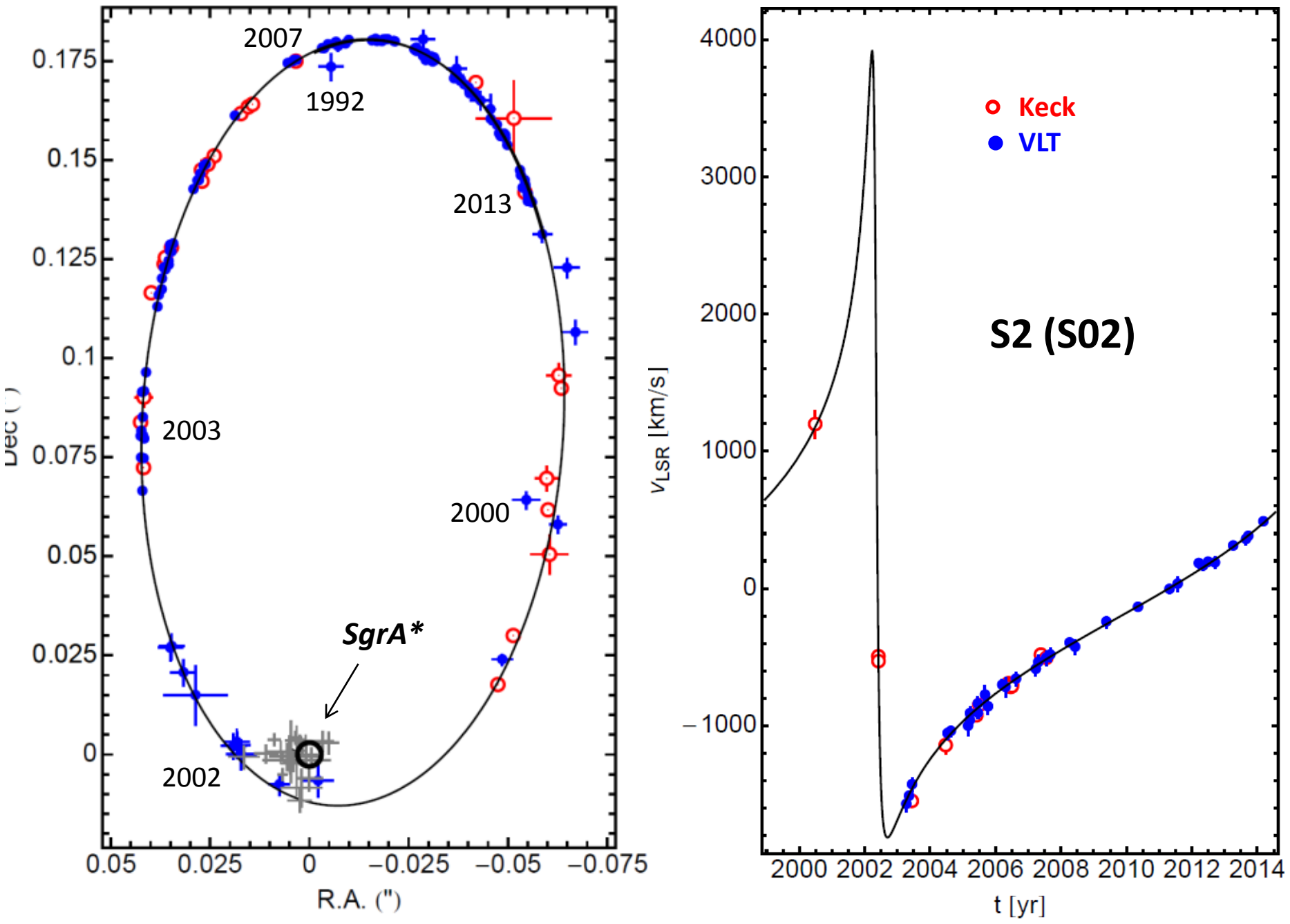

Figure 3. Position on the sky as a function of time (left) and Doppler velocity (relative to the Local Standard of Rest) as a function of time (right) of the star 'S2 (or S02)' orbiting the compact radio source SgrA*. Blue filled circles denote data taken with the ESO NTT and VLT as part of the MPE Galactic Center monitoring project (Schödel et al. 2002, 2005, Gillessen et al. 2009), and red open circles denote data taken with the Keck telescope as part of the UCLA monitoring project (Ghez et al. 2003, 2008, see Gillessen et al. 2009a for the calibration to a common reference frame). Superposed is the best fitting elliptical orbit (continuous curve: central mass 4.26 $(\pm 0.14)_{\text{statistical}}$ $(\pm 0.2)_{\text{systematics}}$ million solar masses, distance 8.36 $(\pm 0.1)_{\text{stat}}$ $(\pm 0.15)_{\text{syst}}$ kpc) with its focus at (0,0) in the left inset (including the independent distance constraints of Reid et al. 2014, Chatzopoulos et al. 2014). The astrometric position of SgrA* is denoted by a circle, grey crosses mark the locations of infrared flares (of typical duration 1-3 hours) that are believed to originate from within the immediate vicinity of the event horizon. The radio source is coincident within the 2 milli-arcsecond errors with the gravitational centroid of the stellar orbit. Since the beginning of the MPE monitoring project (1991/1992), the star has completed its first full orbit in 2007, and it passed its peri-center position 17 light hours from SgrA* in spring 2002 (and again in spring 2018).

VLBI observations have set an upper limit of about 20 km/s and 2 km/s to the motion of SgrA* itself, along and perpendicular to the plane of the Milky Way, respectively (Reid & Brunthaler 2004). When compared to the two orders of magnitude greater velocities of the stars in the immediate vicinity of SgrA*, this demonstrates that the radio source must indeed be massive, with simulations giving a lower limit to the mass of SgrA* of ~$10^5$ solar masses (Chatterjee, Hernquist & Loeb 2002). The intrinsic size of the radio source at about 1mm is only about 4 times the event horizon diameter of a 4 million solar mass black hole (Bower et al. 2004, Shen et al. 2005, Doeleman et al. 2008).

Combining radio size and proper motion limit of SgrA* with the dynamical measurements of the nearby orbiting stars leads to the conclusion that SgrA* can only be a massive black hole, beyond any reasonable doubt (Genzel et al. 2010).

The current Galactic Center evidence eliminates all plausible astrophysical plausible alternatives to a massive black hole. These include astrophysical clusters of neutron stars, stellar black holes, brown dwarfs and stellar remnants (e.g., Maoz 1995, 1998; Genzel et al. 1997, 2000; Ghez et al. 1998, 2005), and even fermion balls (Viollier, Trautmann & Tupper 1993, Munyaneza, Tsiklauri & Viollier 1998, Ghez et al. 2005; Genzel, Eisenhauer & Gillessen 2010). Clusters of a very large number of mini-black holes and boson balls (Torres, Capozziello & Lambiase 2000; Schunck & Mielke 2003; Liebling & Palenzuela 2012) are harder to exclude. The former have a large relaxation and collapse time, the latter have no hard surfaces that could exclude them from luminosity arguments (Broderick, Loeb & Narayan 2009), and they are consistent with the dynamical mass and size constraints. However, such a boson 'star' would be unstable to collapse to a MBH when continuously accreting baryons (as in the Galactic Center), and it is very unclear how it could have formed. Under the assumption of the validity of General Relativity the Galactic Center is now the best quantitative evidence that MBH do indeed exist.

## 5. Massive Black Holes in the local Universe

Beyond the "gold standards" in the Galactic Center and NGC 4258, evidence for the presence of central mass concentrations (which we will henceforth assume to be MBH even though this conclusion can be challenged in most of the individual cases), and a census of their abundance and mass spectrum comes from a number of independent methods,

- robust evidence for MBH in about 10 galaxies comes from VLBI studies of $H_2O$ maser spots in circum-nuclear Keplerian disks of megamaser galaxies akin to NGC4258 (the NRAO "megamaser cosmology project", https://safe.nrao.edu/wiki/bin/view/Main/MegamaserCosmologyProject, Braatz et al. 2010, Kuo et al. 2011, Reid et al. 2013);
- robust evidence for MBH for about 80 galaxies comes from modeling of the spatially resolved, line-of-sight integrated stellar Doppler-velocity distributions with the Hubble Space Telescope (HST) and large ground based telescopes with AO (see the recent reviews of Kormendy & Ho 2013, McConnell & Ma 2013 and references therein). Among the latter, a particularly impressive case is the nucleus of M31, the Andromeda galaxy, where a $10^8$ $M_\odot$ central mass is identified from the rapid (~900 km/s) rotation of a compact circum-nuclear stellar disk (Bender et al. 2005);
- for a number of galaxies, observations of the spatially resolved motions of ionized gas also provide valuable evidence for central mass concentrations, which, however, can be challenged, as mentioned for the Galactic Center, by the possibility of non-gravitational motions (Macchetto et al. 1997, van der Marel & van den Bosch 1998, Barth et al. 2001, Marconi et al. 2003, 2006, Neumayer et al. 2006);

- most recently, high resolution interferometric observations of CO emission have become available as a promising new tool for determining robust central masses (Davis et al. 2013);
- qualitative evidence for the presence of accreting black holes naturally comes for all bona-fide AGN from their IR-optical-UV- and X-ray spectral signatures. In the case of type 1 AGN with broad permitted lines coming from the central light days to light years around the black hole (c.f. Netzer 2013 and references therein), it is possible to derive the size of the broad line region (BLR) from correlating the time variability of the (extended) BLR line emission with that of the (compact) ionizing UV continuum. This reverberation technique (Blandford & McKee 1982) has been successfully applied to derive the BLR sizes (Peterson 1993, 2003, Netzer & Peterson 1997, Kaspi et al. 2000) in several dozen AGN, and has yielded spatially resolved imaging of the BLR in a few (e.g. Bentz et al. 2011). Kaspi et al. (2000). These observations show that the size of the BLR is correlated with the AGN optical luminosity, $R_{BLR} \sim [(\nu L_\nu)_{5100\,Å}]^{0.7}$. After empirical calibration of the zero points of the correlation measurements of the line width of the BLR and of the rest frame optical luminosity of the AGN are sufficient to make an estimate of the MBH mass. As this requires only spectro-photometric data, the technique can be applied even for distant (high redshift) type 1 AGNs (Vestergaard 2004, Netzer et al. 2006, Traktenbrot & Netzer 2012), as well as for low-luminosity AGN in late type and dwarf galaxies (Filippenko & Sargent 1989, Ho, Filippenko & Sargent 1997, Greene & Ho 2004, 2007, Ho 2008, Reines et al. 2011, Greene 2012, Reines, Greene & Geha 2013).

## 6. Demographics and MBH-galaxy “co-evolution”

These data give a fairly detailed census of the incidence and of the mass spectrum of the local (and less so, also of the distant) MBH population. MBH masses span a range at least five orders of magnitudes from $10^5\ M_\odot$ in dwarf galaxies to $10^{10}\ M_\odot$ in the most massive central cluster galaxies. Most massive spheroidal/bulged galaxies appear to have a central MBH. The occupation fraction drops in bulgeless systems with decreasing galaxy mass (Greene 2012). It is not clear yet whether the lack of observational evidence below $10^5\ M_\odot$ is real, or driven by observational detectability. The inferred black hole mass and the mass of the galaxy’s spheroidal component (but not its disk, or dark matter halo) are strongly correlated (Magorrian et al. 1998, Häring & Rix 2004). The most recent analyses of Kormendy & Ho (2013) and McConnell & Ma (2013) find that between 0.3 and 0.5% of the bulge/spheroid mass is in the central MBH. The scatter of this relation is between ±0.3 and ±0.5 dex, depending on sample and analysis method (McConnell & Ma 2013). A correlation of comparable scatter exists between the black hole mass and the bulge/spheroid velocity dispersion σ ($M_{BH} \sim \sigma^\beta$, with β~4.2-5.5, Ferrarese & Merritt 2000, Gebhardt et al. 2000, Tremaine et al. 2002, Kormendy & Ho 2013, McConnell & Ma 2013, Figure 4).

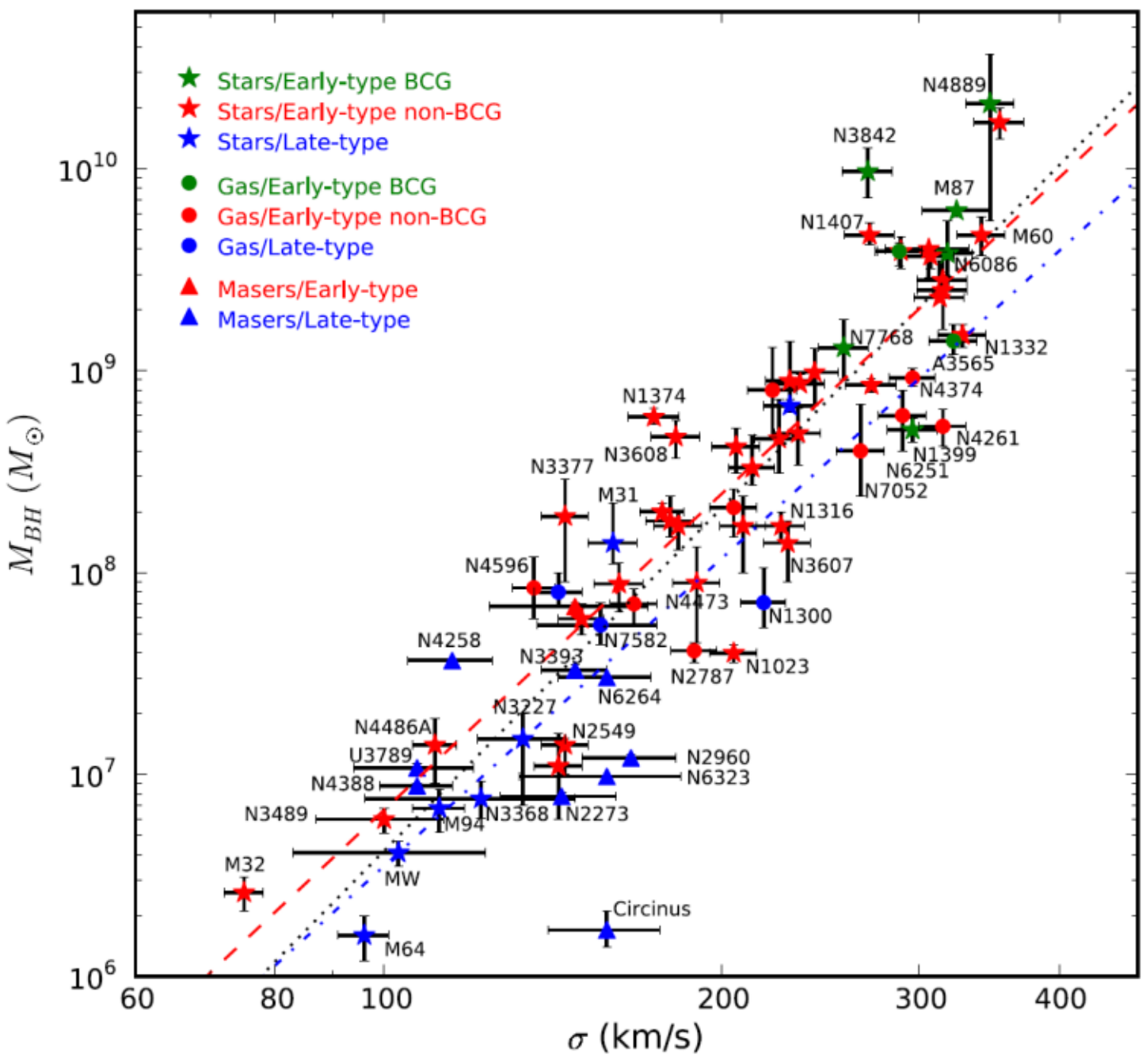

Figure 4. Black hole mass $M_{BH}$ (vertical axis) as a function of galaxy velocity dispersion σ (horizontal axis), for the all 72 galaxies in the compendium of McConnell & Ma (2013). Asterisks, filled circles and filled triangles denote the technique that was used to determine the MBH mass (stellar kinematics, gas kinematics, or masers), and red, green and blue colors denote the type of host galaxy (spheroidal galaxy, very massive spheroidal galaxy at the center of a galaxy cluster (BCG), and late type (disk/ irregular) galaxy). The black dotted line shows the best-fitting power law for the entire sample: log10 ($M_{BH}/M_\odot$) = 8.32 + 5.64 log(σ/200km/s). When early-type and late-type galaxies are fitted separately, the resulting power laws are log($M_{BH}/M_\odot$) = 8.39 + 5.20 log(σ/200km/s) for the early-type (red dashed line), and log ($M_{BH}/M_\odot$) = 8.07 + 5.06 log (σ/200km/s) for the late-type galaxies (blue dot-dashed line). The plotted values of σ are derived using kinematic data within the effective radius of the spheroidal galaxy component (adapted from McConnell & Ma 2013).

Ever since this correlation between central black hole mass and galaxy host spheroidal mass (or velocity dispersion) component has been established, the interpretation has been that there must be an underlying connection between the formation paths of the galaxies' stellar components and their embedded central MBHs. This underlying connection points back to the peak formation epoch of massive galaxies about 6-10 Gyrs ago (e.g. Madau et al. 1996, Haehnelt 2004). The fact that the correlation is between the black hole mass and the bulge/spheroidal component, and not the total galaxy or dark matter mass, has been taken as evidence that most of the MBH's growth, following an early evolution from a lower mass seed, is triggered by a violent dissipative process at this early epoch. The most obvious candidate are major mergers between early gas rich galaxies, which are widely thought to form bulges in the process (Barnes & Hernquist 1996, Kauffmann & Haehnelt 2000, Haiman & Quataert 2004, Hopkins et al. 2006, Heckman et al. 2004). Compelling support for the AGN – merger model comes from the empirical evidence that dusty ultra-luminous infrared galaxies (ULIRGs, $L_{IR}>10^{12}L_{\odot}$) in the local Universe are invariably major mergers of gas-rich disk galaxies (Sanders et al. 1988); the majority of the most luminous late stage ULRGs are powered by obscured AGN (Veilleux et al. 1999, 2009).

This 'strong' co-evolution model is further supported by the fact that the peak of cosmic star formation 10 Gyrs ago is approximately coeval with the peak of cosmic QSO activity (Boyle et al. 2000), and that the amount of radiation produced during this QSO era is consistent with the mass present in MBHs locally for a 10-20% radiation efficiency during MBH mass growth (Soltan 1982, Yu et al. 2002, Marconi et al. 2004, Shankar et al. 2009). There is an intense ongoing discussion whether or not MBHs and their hosts galaxies formed coevally and grew on average in lock-step (Figure 5, Marconi et al. 2004, Shankar et al. 2009, Alexander & Hickox 2012, Mullaney et al. 2012, del Vecchio et al. 2014), or whether MBHs started slightly earlier or grew more efficiently (Jahnke et al. 2009, Merloni et al. 2010, Bennett et al. 2011). The fact that the correlation appears to be quite tight suggests that feedback between the accreting and rapidly growing black holes during that era and the host galaxy may have been an important contributor to the universal shutdown of star formation and mass growth in galaxies above the Schechter mass, $M_S \geq 10^{10.9}$ $M_{\odot}$ (Baldry et al. 2008, Conroy & Wechsler 2009, Peng et al. 2010, Moster et al. 2013, Behroozi et al. 2013).

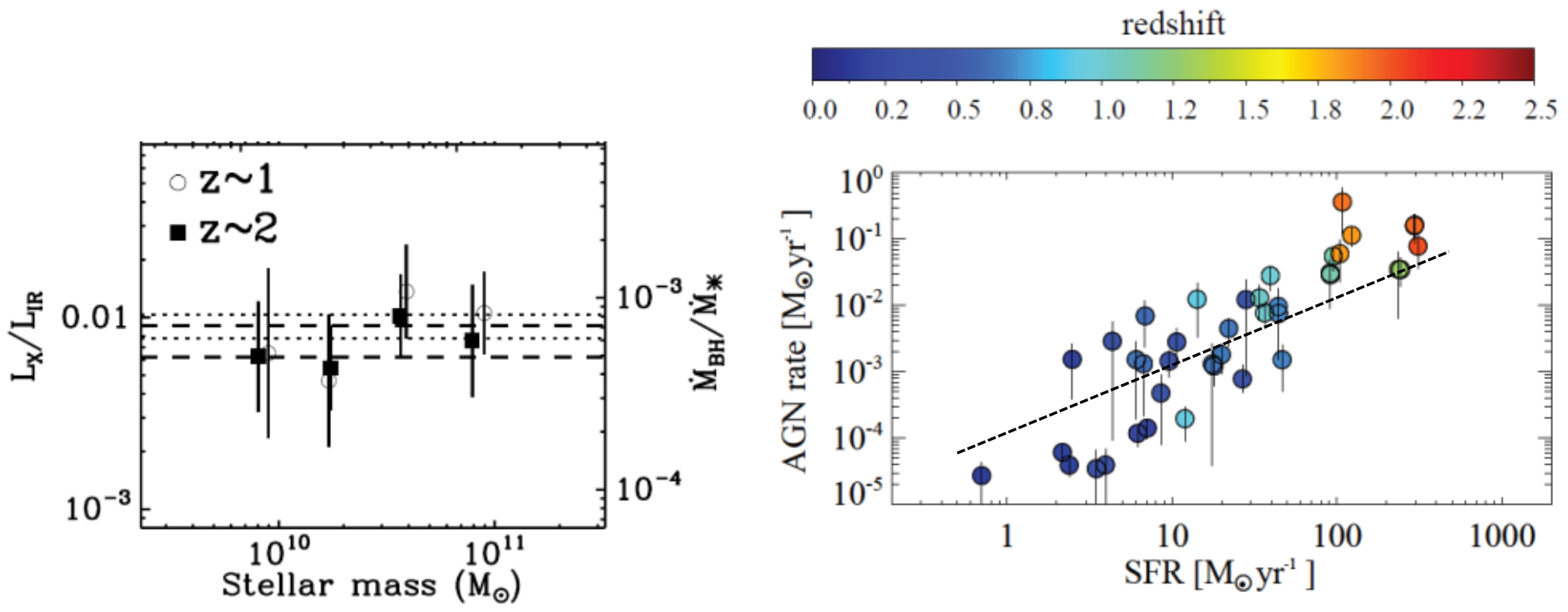

Figure 5. Evidence for average MBH-galaxy growth co-evolution by stacking deep X-ray data (as quantitative indicators of AGN growth) on multi-wavelength images of star forming galaxies (including mid- and far-IR emission as extinction- and (nearly) AGN-independent tracers of star formation rates) in GOODS-S (left, Mullaney et al. 2012) and GOODS-S and COSMOS (right, delVecchio et al. 2014). The left plot shows that the ratio of inferred black hole growth to star formation rate is the same in several mass bins and at z~1 and z~2. The right plot shows the integrated MBH growth rate as a function of star formation and redshift (colors). The dotted line has slope unity and is not a fit to the data. However, once the dependence on mass and redshift are de-coupled, the best fitting correlation does have unity slope, suggesting average co-evolution.

## 7. AGN-MBH feedback

Throughout the last 10 billion years galaxies have been fairly inefficient in incorporating the cosmic baryons available to them into their stellar components. At a dark matter halo mass near $10^{12}$ $M_\odot$ this baryon fraction is only about 20% (of the cosmic baryon abundance), and the efficiency drops to even lower values on either side of this mass (e.g., Madau et al. 1996; Baldry et al. 2008, Conroy & Wechsler 2009, Guo et al. 2010, Moster et al. 2013, Behroozi et al. 2013). Galactic winds driven by supernovae and massive stars have long been proposed to explain the low baryon content of halos much below $\log(M_h/M_\odot)$~12 (e.g. Dekel & Silk 1986, Efstathiou 2000). The decreasing efficiency of galaxy formation above $\log(M_h/M_\odot)$~12 may be caused by less efficient cooling and accretion of baryons in massive halos (Rees & Ostriker 1977, Dekel & Birnboim 2006). Alternatively or additionally efficient outflows driven by accreting MBH may quench star formation at the high mass tail, at and above the Schechter stellar mass, $M_S$~$10^{10.9}$ $M_\odot$ (di Matteo, Springel & Hernquist 2005, Croton et al. 2006, Bower et al. 2006, Hopkins et al. 2006, Cattaneo et al. 2007, Somerville et al. 2008, Fabian 2012).

In the local Universe, such ‘AGN or MBH feedback’ has been observed in the so called ‘radio mode’ in very massive, central cluster galaxies driving jets into the intra-cluster medium. In these cases the central MBHs typically is in a fairly low or quiescent radiative state. Considerations of energetics suggest that radio mode feedback plausibly prevents cooling cluster gas to fall onto these massive galaxies that would otherwise lead to substantial further star formation and mass growth (McNamara & Nulsen 2007, Fabian 2012, Heckman & Best 2014). A second MBH feedback mode (termed ‘QSO mode’), in which the MBH is active (i.e. the AGN is luminous) is detected as ionized winds from AGN (e.g. Cecil, Bland, & Tully 1990, Veilleux, Cecil & Bland-Hawthorn 2005, Westmoquette et al. 2012, Rupke & Veilleux 2013, Harrison et al. 2014) and from obscured QSOs (Zakamska & Greene 2014). The QSO mode feedback in form of powerful neutral and ionized gas outflows has also been found in late stage, gas rich mergers (Fischer et al. 2010, Feruglio et al. 2010, Sturm et al. 2011, Rupke & Veilleux 2013, Veilleux et al. 2013), which however are rare in the local Universe.

At high-z AGN QSO mode feedback has been seen in broad absorption line quasars (Arav et al. 2001, 2008, 2013, Korista et al. 2008), in type 2 AGN (Alexander et al. 2010, Nesvadba et al. 2011, Cano Díaz et al. 2012, Harrison et al. 2012), and in radio galaxies (Nesvadba et al.2008). However, luminous AGNs near the Eddington limit are again rare, constituting less than 1% of the star forming population in the same mass range (e.g. Boyle et al. 2000). QSOs have short lifetimes relative to the Hubble time ($t_{QSO}$~ $10^7$ – $10^8$ yr << $t_H$, Martini 2004) and thus have low duty cycles compared to galactic star formation processes ($t_{SF}$~ $10^9$ yr, Hickox et al. 2014). It is thus not clear whether the radiatively efficient QSO mode can have much effect in regulating galaxy growth and star formation shutdown, as postulated in the theoretical work cited above (Heckman 2010, Fabian 2012).

From deep adaptive optics assisted integral field spectroscopy at the ESO VLT, Förster Schreiber et al. (2014) and Genzel et al. (2014) have recently reported the discovery of broad (~$10^3$ km/s), spatially resolved (a few kpc) ionized gas emission associated with the nuclear regions of very massive ($\log(M_*/M_\odot)$>10.9) z~1-2 star forming galaxies (SFGs). While active AGN do exhibit similar outflows, as stated above, the key breakthrough of this study is that it provides compelling evidence for wide-spread and powerful nuclear outflows in most (~70%) normal massive star forming galaxies at the peak of galaxy formation activity. The fraction of active, luminous AGN among this sample is 10-30%, suggesting that the nuclear outflow phenomenon has a significantly higher duty cycle than the AGN activity. If so, MBHs may indeed be capable to contribute to the quenching of star formation near the Schechter mass, as proposed by the theoretical work mentioned above.

## 8. Non-Merger Evolution Paths of MBHs

The most recent data on MBH demographics (Figure 4) suggest that the simple scenario of early MBH-galaxy formation through mergers and strong "co-evolution" might be too simplistic. Kormendy & Ho (2013, see also Kormendy, Bender & Cornell 2011) as well as McConnell & Ma (2013) find that MBHs in late type galaxies tend to fall below the best correlation of the pure spheroidal systems. The "pseudo"-bulges in these disk galaxies (including the Milky Way itself) typically rotate rapidly and may have partially formed by radial transport of disk stars to the nucleus mediated through slow, secular angular momentum transport, rather than by rapid merger events. In these systems the efficiency and growth processes of MBHs appears to be lower than in the very massive spheroids that formed a long time ago. In the local Universe, the Sloan Digital Sky Survey has shown that most AGN are not involved in active mergers or galaxy interactions (Li et al. 2008). Most lower luminosity AGN are in massive early type hosts that are not actively fed. Most of the lower-mass MBH growth at low redshift happens in lower mass galaxies (Kauffmann et al. 2003, Heckman et al. 2004).

Lower mass (<$10^{5.3..7}$ $M_\odot$) MBHs have been found in bulge-less disks and even dwarf galaxies (Filippenko & Ho 2003, Barth et al. 2001, Barth, Greene & Ho 2005, Greene & Ho 2004, 2007, Reines et al. 2011, 2013), in which there appears to be no or little correlation between the properties of the galaxy and its central MBH, in contrast to the bulged/spheroid systems (Greene 2012). These MBHs must have formed more through an entirely different path. MBH growth in these cases is more likely to be controlled by local processes, such gas infall from local molecular clouds (Sanders 1998, Genzel et al. 2010 and references therein) and stellar mass loss following a nuclear 'starburst' (Scoville & Norman 1988, Heckman et al.2004, Davies et al. 2007, Wild et al. 2010).

At the peak of the galaxy formation epoch (redshifts z~1-2) imaging studies show little evidence for the average AGNs to be in ongoing mergers (Cisternas et al. 2011, Schawinski et al. 2011, Kocevski et al. 2012). Instead most AGNs at this epoch are active star forming galaxies, including large disks, near the 'main-sequence' of star formation (Shao et al. 2010, Rosario et al. 2012, 2013). For active MBH, AGN luminosity and star formation rates are not or poorly correlated, excepting at the most extreme AGN luminosities (Netzer 2009, Rosario et al. 2012), yet the average MBH and galaxy growth rates are (Mullaney et al. 2012, del Vecchio et al. 2014). The empirical evidence for the AGN-merger model based on luminosity functions and spatial correlations (Hopkins et al. 2006) has been shown to not be a unique interpretation (Conroy & White 2013).

All these findings suggest that the concept of co-evolution between MBH growth and galaxy growth may most of the time be applicable only on average, or merely as a non-causal, statistical 'central limit' (Jahnke & Maccio 2011). One might call this 'weak' co-evolution. The instantaneous MBH growth rate at any given time exhibits large amplitude fluctuations (Hopkins et al.2005, Novak et al. 2011, Rosario et al. 2012, Hickox et al. 2014). Relatively rare gas rich mergers may be able to stimulate phases of strong co-evolution at all redshifts. At other times, radial transport of gas (and stars) in galaxy disks may be an alternative channel of MBH growth, at least at the peak of galaxy-MBH formation, since galaxies 10 billion years were gas rich (Tacconi et al. 2013), resulting in efficient radial transport from the outer disk to the nucleus (a few hundred million years, Bournaud et al. 2011, Alexander & Hickox 2012). These inferences from the empirical data are in good agreement with the most recent hydrodynamical simulations (Sijacki et al. 2014).

## 9. MBH Spin

X-ray spectroscopy of the 6.4-6.7 keV Fe K-complex finds relativistic Doppler motions in several tens of AGNs, following the initial discovery in the iconic Seyfert galaxy MCG-6-30-15 (Tanaka et al. 1995, Figure 5). The Fe-K profiles can be modelled as a rotating disk on a scale of few to 20 $R_S$ that reflects a power law, hard X-ray continuum emission component likely located above the disk (Tanaka et al. 1995, Nandra et al. 1997, 2007, Fabian et al. 2000, 2002, Fabian & Ross 2010, Reynolds 2013). While the X-ray spectroscopy by itself does not yield black hole masses, it provides strong support for the black hole interpretation. In addition, reverberation techniques of the time variable spectral properties are beginning to deliver interesting constraints on the spatial structure of the continuum and line components (Fabian et al. 2009, Uttley et al. 2014). From the modeling of the spectral profiles it is possible to derive unique constraints on the MBH spin, assuming that the basic modeling assumptions are applicable. The inferred spin for MCG-6-30-15 is near maximal (Figure 6). In a sample of 20 MBHs investigated in this way at least half have a spin parameter a>0.8, providing tantalizing, exciting evidence for a frequent occurrence of high-spin MBH (Figure 6, Reynolds 2013 and references therein). These measurements promise to yield important information on the growth processes of MBH.

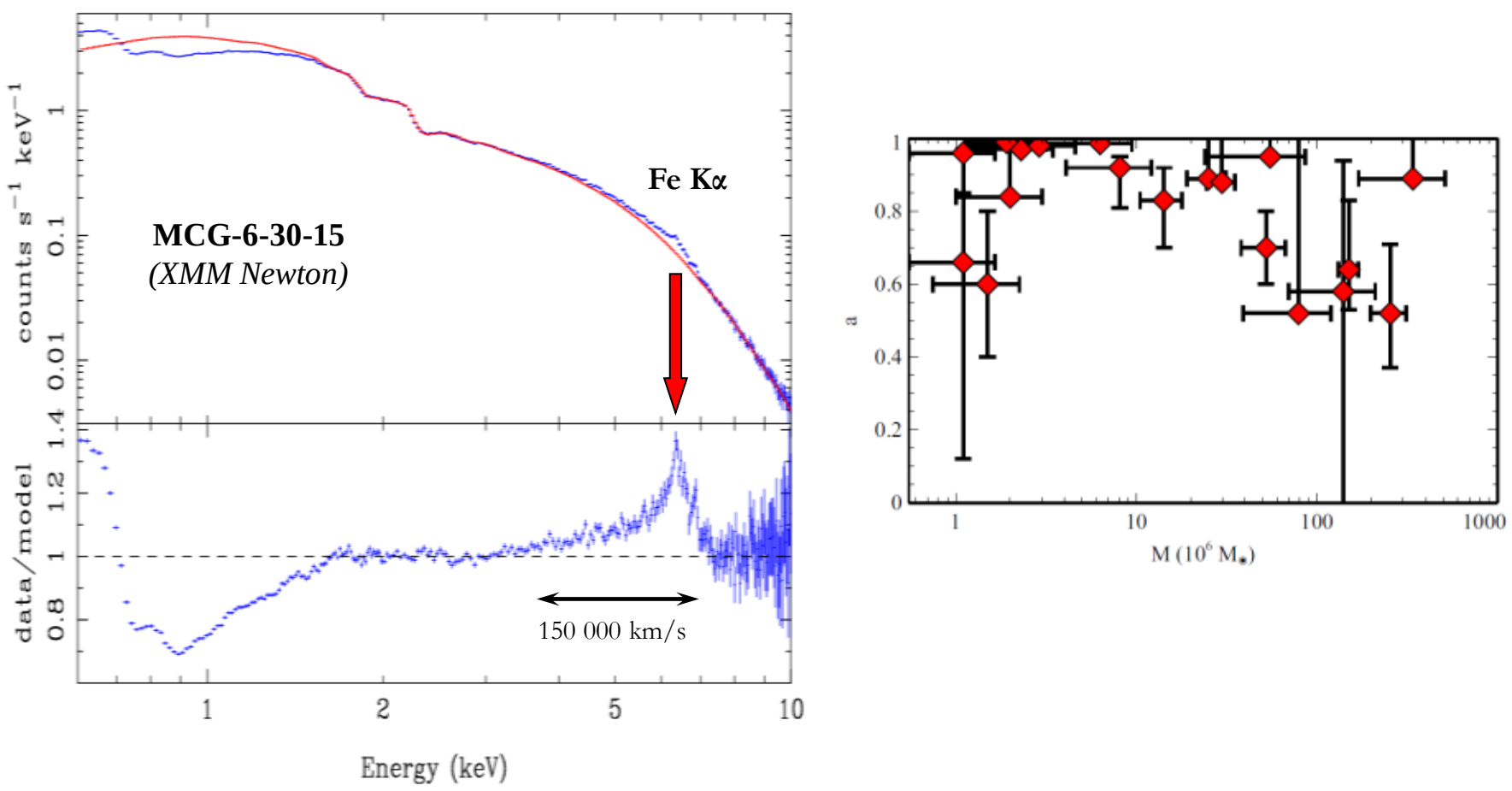

Figure 6. Left: Fe-K profile of the iconic Seyfert galaxy MCG-6-30-15 (Tanaka et al. 1995) obtained with XMM-Newton (Fabian et al. 2002, Fabian & Vaughan 2003). The blue extension of the relativistic emission extends as low as ~ 3 keV. In the framework of a rotating accretion disk reflecting a hard X-ray power law component this means that there is emission at ~ 2×$R_{grav}$. If the MBH has a low angular momentum parameter, this would require significant reflection occurring within the last stable orbit. Alternatively, the MBH in MCG-6-3-15 is a near maximum Kerr hole with a>0.97 (Brenneman & Reynolds 2006). Right: Inferred spin parameter a as a function of black hole mass, for 20 MBHs (Reynolds 2013)

## 10. Early Growth

The formation and evolution of MBHs faces two basic problems. One is angular momentum. To make it into the MBH event horizon from the outer disk of a galaxy, a particle has to lose all but $10^{-9}$ of its original angular momentum, a truly daunting task (c.f. Alexander & Hickox 2012). For this reason, major mergers have been considered natural candidate for being the sites of rapid MBH growth (Hopkins et al. 2006), since the mutual gravitational torques in a galaxy-galaxy interaction can reduce more than 90% of the angular momentum of a significant fraction of the total interstellar gas (e.g. Barnes & Hernquist 1996). However, this is by far not sufficient. It is likely that several stages of additional angular momentum loss much closer to the nucleus are involved in growing MBH, plausibly including star formation events at different 'way-points' including the nucleus itself (Scoville & Norman 1988, Davies et al. 2007, Hopkins & Quataert 2010, Meyer et al. 2010, Wild, Heckman & Charlot 2010, Alexander & Hickox 2012 and references therein). For this reason, black hole growth, accretion and radiation are probably highly time variable and strongly influenced by the properties of the gas and stellar environment in the sphere of influence around the black hole (Genzel et al. 2010, Hickox et al. 2014).

The second major obstacle is the time needed to grow to a final mass M from an initial seed of much lower mass $M_0$ (Volonteri 2010, 2012). This time is given by

$$\frac{t_M}{t_{Salpeter}} = \frac{\eta}{1-\eta} \times \frac{1}{L/L_{Edd}} \ln\left(\frac{M}{M_0}\right) \qquad (1),$$

where $t_{Salpeter}=4\times10^8$ yr, η is the radiative efficiency, $L_{Edd}$ is the Eddington luminosity ($3.4\times10^4$ M ($M_\odot$) where the accreting MBH's radiation pressure equals its gravity. To grow to $M=10^9\ M_\odot$ at z~6 (~1 billion yr after the Big Bang) with η~0.1 at the Eddington rate requires $4\times10^7 \ln(M/M_0)$ yr. If the initial seed formed in the re-ionization epoch at z~10, the seed mass has to exceed $\sim10^4 M_\odot$. While $10^9\ M_\odot$ MBHs as early as z~6 are rare ($10^{-10}$ $Mpc^{-3}$, Fan et al. 2006), and most very massive MBHs could have reached their final masses later, this example does show that standard Eddington accretion from a relative low mass seed, such as a super massive star ($M_0\sim10^2\ M_\odot$), cannot account for the oldest MBHs. Possibilities include fairly massive seeds ($\geq10^4\ M_\odot$) formed from direct collapse of a dense gas cloud (Silk & Rees 1998), perhaps including a phase of super-Eddington accretion (see the more detailed account in Mitch Begelman's contribution in this volume, as well as the discussion in Begelman, Volonteri & Rees 2006, Volonteri 2010, 2012).

## 10. Zooming in on the Event Horizon

Looking forward to the next decade, there are several avenues to get still firmer constraints on the black hole paradigm, and determine the gravitational field still closer to the event horizon, in particular in the Galactic Center. Infrared spectroscopy of S2 during the next peri-approach in 2018 will have a good chance of detecting post-Newtonian parameters (Roemer effect, gravitational redshift, longitudinal Doppler effect, e.g. Zucker et al. 2006). Within the next decade it should be possible with current astrometric capabilities to detect S2's Schwarzschild precession angle, $\Delta\Phi_S = \frac{3\pi}{1-e^2}\left(\frac{R_S}{a}\right) \sim 12'$. The Schwarzschild precession and perhaps even the Lense-Thirring precession (due to the spin and quadrupole moment of the MBH) are obviously more easily detectable for stars with smaller semi-major axes and shorter orbital periods than S2. One such star, S102/S55 has been reported by Meyer et al. (2012), but the current confusion limited imagery on 10m-class telescopes prevents further progress.

This barrier will be broken in the next few years by the GRAVITY near-IR interferometric experiment (Eisenhauer et al. 2005a, 2011, Gillessen et al. 2006), and with the next generation 30m-class telescopes a decade later (Weinberg et al. 2005). GRAVITY will combine the four VLT telescopes interferometrically, with the goal of 10μarcsec precision, near-infrared imaging interferometry (angular resolution a few milli-arcseconds). GRAVITY will also be able to search for dynamical signatures of the variable infrared emission from SgrA* itself (Genzel et al. 2003, Eckart et al. 2006, Do et al. 2009, Dodds-Eden et al. 2009, c.f. Baganoff et al. 2001). These 'flares' originate from within a few milli-arcseconds of the radio position of SgrA* and probably occur when relativistic electrons in the innermost accretion zone of the black hole are substantially accelerated so that they are able to produce infrared synchrotron emission and X-ray synchrotron or inverse Compton radiation (Markoff et al. 2001). As such the infrared variable emission as well as the millimeter and submillimeter emission from SgrA* probe the inner accretion zone between a few to 100 $R_S$. If orbital motion (of hot spots) could be detected the space time very close to the event horizon could potentially be probed (Paumard et al. 2005, Broderick & Loeb 2006, 2009, Hamaus et al. 2009).

VLBI at short millimeter or submillimeter wavelengths may be able to map out the strong light bending ('shadow') region inside the photon orbit of the MBH (Bardeen 1973, Falcke, Melia & Algol 2000). This 'Event Horizon Telescope' Project ( http://www.eventhorizontelescope.org/) will soon benefit the observations of the Galactic Center (and the relatively nearby ~$6x10^9$ $M_\odot$ MBH in M87) from the much enhanced sensitivity and additional u-v-coverage with the ALMA interferometer in Chile (Lu et al. 2014). It is hoped that the shadow signature can be extracted fairly easily even from data with a sparse coverage of the UV-plane (e.g. Doeleman 2010). As in the case of the GRAVITY observations of the infrared flares, it is not clear, however, how the potentially complex emission structure from the inner accretion zone, including a possible radio jet, may compromise the interpretation of EHT maps in terms of GR effects (Dexter & Fragile 2013,Moscibrodzka et al. 2014).

Given the current presence of ~200 OB stars in the central parsec (Genzel et al. 2010) and extrapolating to earlier star formation episodes, there should be 100-1000 neutron stars, and thus potentially many pulsars within the parsec-scale sphere of influence of the Galactic Center MBH (Pfahl & Loeb 2004, Wharton et al. 2012). Until recently none have been found, despite many radio searches. The blame was placed on the large dispersion of the radio pulses by large columns of electrons in front of the Galactic Center sight line. In 2013 SWIFT and NUSTAR discovered a magnetar, 1745-2900, within 3" of SgrA*, with a pulse period of 3.7 s, whose radio pulse characteristics have since been studied in detail (Kennea et al. 2013, Mori et al. 2013, Eatough et al. 2013, Spitler et al. 2014, Bower et al. 2014). While the magnetar itself cannot be used for timing studies, its detection renews the hope that radio pulsars can be detected with sufficient sensitivity but also increases the suspicion that there are physical germane to the Galactic Center region suppressing the formation of normal radio pulsars. If radio pulsars can be detected, however, precision timing, especially with the future capabilities of the Square Kilometer Array (SKA), has the potential to detect not only post-Newtonian parameters (including the Shapiro delay and the Schwarzschild precession term), but also the Lense-Thirring and quadrupole terms (Liu et al. 2012).

If one or several of these efforts is successful, it may be ultimately possible to test GR in the strong curvature limit, and test the no-hair theorem (Will 2008, Merritt et al. 2010, Psaltis & Johanssen 2011).

*Acknowledgements: I would like to thank Stefan Gillessen, David Rosario, Amiel Sternberg, Scott Tremaine and Stijn Wuyts for helpful comments on this manuscript.*

## References

Alexander, D. M., Swinbank, A. M., Smail, I., McDermid, R., & Nesvadba, N. P. H. 2010, MNRAS, 402, 2211

Alexander, D.M. & Hickox, R.C., 2012, New.Astr.Rev., 56, 92

Almheiri, A. Marolf, D., Polchinski, J. & Sully, J. 2013. JHEP 02. 62

Arav, N., de Kool, M., Korista, K. T., et al. 2001, ApJ, 561, 118

Arav, N., Moe, M., Costantini, E., Korista, K. T., Benn, C. & Ellison, S. 2008, ApJ, 681, 954

Arav, N., Borguet, B., Chamberlain, C., Edmonds, D. & Danforth, C. 2013, MNRAS, 436, 3286

Argon, A.L., Greenhill, L.J., Reid, M.J., Moran, J.M. & Humphreys, E.M.L. 2007, ApJ, 659, 1040

Baganoff, F. et al.2001, Nature, 413, 45

Baldry, I. K., Glazebrook, K., & Driver, S. P. 2008, *MNRAS* 388, 945

Bardeen J. M., 1973, in Black holes (Les astres occlus), DeWitt B. S., DeWitt C., eds., New York: Gordon and Breach, p. 215

Barnes, J.E. & Hernquist, L. 1996, ApJ, 471, 115

Barth, A.J. Sarzi, M., Rix, H.-W., Ho, L. C., Filippenko, A.V. & Sargent, W. L. W. 2001, ApJ, 555, 685

Barth, A.J., Greene, J.E. & Ho, L.C. 2005, ApJ, 619, L151

Begelman, M.C., Volonteri, M. & Rees, M.J. 2006, MNRAS, 370, 289

Behroozi, P.S., Wechsler, R.H. & Conroy, C. 2013, ApJ 770, 57

Bender, R. et al. 2005, ApJ 631, 280

Bennert, V. N., Auger, M.W., Treu, T., Woo, J.-H.& Malkan, M.A. 2011, ApJ, 742, 107

Bentz, M.C. et al. 2010, ApJ, 720, L46

Blandford, R.D. & McKee, C.F. 1982, ApJ, 255, 419

Blandford, R.D. 1999, ASPC, 160, 265

Bournaud, F., Dekel, A., Teyssier, R., Cacciato, M., Daddi,E., Juneau, S. & Shankar, F. 2011, ApJ, 741, L33
Bousso, R. 2002, Rev.Mod.Phys. 74, 825
Bower, G.C. et al. 2004, Science 304, 704
Bower, G.C. et al. 2014, ApJ, 780, L2
Bower, R.G, Benson, A. J., Malbon, R., Helly, J. C., Frenk, C. S., Baugh, C. M., Cole, S. & Lacey, C. G. 2006, MNRAS, 370, 645
Boyle, B. J.,Shanks, T.,Croom, S. M.,Smith, R. J.,Miller, L.,Loaring, N. & Heymans, C. 2000, MNRAS, 317, 1014
Braatz, J.A., Reid, M.J., Humphreys, E.M.L., Henkel, C., Condon, J.J. & Lo, K.Y. 2010, ApJ, 718, 657
Brenneman, L.W. & Reynolds, C.S. 2006, ApJ, 652, 1028
Broderick, A. E. & Loeb, A. 2006, MNRAS, 367, 905
Broderick, A., Loeb, A. & Narayan, R. 2009, ApJ, 701, 1357
Broderick, A. E., Fish, V. L.; Doeleman, S. S.& Loeb, A. 2009, ApJ, 697, 45
Cano-Díaz, M., Maiolino, R., Marconi, A., et al. 2012, A&A, 537, L8
Cattaneo, A., Dekel, A., Devriendt, J., Guiderdoni, B.& Blaizot, J. 2006, MNRAS, 370, 1651
Cecil, G., Bland, J. & Tully, B. R. 1990, ApJ, 355, 70
Chatterjee, P., Hernquist, L. & Loeb, A. 2002, ApJ, 572, 371
Chatzopoulos, S., Fritz, T., Gerhard, O., Gillessen, S., Wegg, C., Genzel, R. & Pfuhl, O. 2014, arXiv1403.5266
Cisternas, M. et al. 2011, ApJ, 726, 57
Conroy, C. & Wechsler, R.H. 2009, ApJ, 696, 620
Conroy, C. & White, M. 2013, ApJ, 762, 70
Croton, D.J. et al. 2006, Mon.Not.Roy.Astr.Soc. 365, 11
Davies, R.I., Müller Sánchez, F., Genzel, R., Tacconi, L. J., Hicks, E. K. S., Friedrich, S.& Sternberg, A. 2007, ApJ, 671, 1388
Davis, T. A., Bureau, M., Cappellari, M., Sarzi, M. & Blitz, L. 2013, Nature, 494, 328
Dekel, A. & Silk, J. 1986, ApJ 303, 39
Dekel, A. & Birnboim, Y. 2006, MNRAS 368, 2
DelVecchio, I. Lutz, D., Berta, S. et al. 2014, in prep.
Dexter, J. & Fragile, P.C. 2013, MNRAS, 432, 2252
Di Matteo, T., Springel, V. & Hernquist, L. 2005, Nature, 433, 604
Do, T., Ghez, A. M., Morris, M. R., Yelda, S., Meyer, L., Lu, J. R., Hornstein, S. D.& Matthews, K. 2009, ApJ, 691, 1021
Dodds-Eden, K. et al. 2009, ApJ, 698, 676
Doeleman, S.S. et al. 2008, Nature, 455, 78
Doeleman, S.S. 2010, in Proceedings of the 10th European VLBI Network Symposium and EVN Users Meeting: VLBI and the new generation of radio arrays. September 20-24, 2010. Manchester, UK. Published online at [http://pos.sissa.it/cgi-bin/reader/conf.cgi?confid=125](http://pos.sissa.it/cgi-bin/reader/conf.cgi?confid=125), id.53
Eatough, R.P. et al. 2013, Nature, 501, 391
Eckart, A. & Genzel, R. 1996, Nature 383, 415
Eckart, A. & Genzel, R. 1997, MNRAS, 284, 576
Eckart, A. et al., 2006, A&A 450, 535

Efstathiou, G. 2000, MNRAS, 317, 697
Einstein, A. 1916, Ann.Phys. 49, 50
Eisenhauer, F. et al. 2003, ApJ 597, L121
Eisenhauer, F. et al. 2005, ApJ 628, 246
Eisenhauer, F., Perrin, G., Rabien, S., Eckart, A., Lena, P., Genzel, R., Abuter, R., & Paumard, T. 2005a, Astr.Nachrichten 326, 561
Eisenhauer, F. et al. 2011, ESO Msngr.143, 16
Fabian, A.C., Iwasawa, K., Reynolds, C.S. & Young, A.I., 2000, PASP, 112, 1145
Fabian, A. C., Vaughan, S., Nandra, K., Iwasawa, K., Ballantyne, D. R., Lee, J. C., De Rosa, A., Turner, A.& Young, A. J. 2002, MNRAS, 335, L1
Fabian, A.C. & Vaughan, S. 2003, MNRAS 340, L28
Fabian, A.C. et al. 2009, Nature, 459, 540
Fabian, A.C. & Ross, R.R., 2010, Sp. Sc. Rev. 157, 167
Fabian, A. C. 2012, ARA&A, 50, 455
Falcke, H., Melia, F. & Algol, E. 2000, ApJ 528, L13
Fan, X. et al. 2006, AJ, 132, 117
Feruglio, C., Maiolino, R., Piconcelli, E., et al. 2010, A&A, 518, L155
Ferrarese, L. & Merritt, D. 2000, ApJ 539, L9
Filippenko, A.V. & Sargent, W.L.W. 1989, ApJ, 342, L11
Filippenko, A.V. & Ho, L.C. 2003, ApJ, 588, L13
Fischer, J., Sturm, E., González-Alfonso, E., et al. 2010, A&A 518, L41
Förster Schreiber, N. M., Genzel, R., Newman, S. F., et al. 2014, ApJ, 787, 38
Gebhardt, K. et al. 2000, ApJ, 539, L13
Genzel, R., Hollenbach, D., & Townes, C. H., 1994, Rep. Prog. Phys., 57, 417
Genzel, R., Thatte, N., Krabbe, A., Kroker, H. & Tacconi-Garman, L.E. 1996, ApJ, 472, 153
Genzel, R., Eckart, A., Ott, T. & Eisenhauer, F. 1997, MNRAS, 291, 219
Genzel, R., Pichon, C., Eckart, A., Gerhard, O.E. & Ott, T. 2000, MNRAS, 317, 348
Genzel, R. et al. 2003, Nature, 425, 934
Genzel, R., Eisenhauer, F. & Gillessen, S. 2010, Rev.Mod.Phys. 82, 3121
Genzel, R. et al. 2014, ApJ in press (arXiv1406.0183)
Ghez, A.M., Klein, B.L., Morris, M. & Becklin, E.E. 1998, ApJ 509, 678
Ghez, A.M. et al. 2000, Nature, 407, 349
Ghez, A.M. et al. 2003, ApJ 586, L127
Ghez, A. et al. 2005, ApJ, 620, 744
Ghez, A. et al. 2008, ApJ, 689, 1044
Giacconi, R., Gursky, H., Paolini, F. & Rossi, B.B. 1962, Phys.Rev.Lett. 9, 439
Giacconi, R. 2003, Rev.Mod.Phys. 75, 995
Gillessen, S. 2006, SPIE, 6268, E11
Gillessen, S. et al. 2009, ApJ, 692, 1075
Gillessen, S. et al. 2009a, ApJ 707, L114
Greene, J.E. & Ho, L.C. 2004, ApJ, 610, 722
Greene, J.E. & Ho, L.C. 2007, ApJ 667, 131
Greene, J.E., 2012, Nature Comm., 10.1038, 2314
Gültekin, K. et al. 2009, ApJ 698, 198
Guo, Q., White, S., Li, C., & Boylan-Kolchin, M. 2010, *MNRAS* 404, 1111

Haehnelt, M. 2004 in ‘Coevolution of Black Holes and Galaxies’, Carnegie Observatories Centennial Symposia. Cambridge University Press, Ed. L.C. Ho, p.405
Häring, N. & Rix, H.-W. 2004, ApJ, 604, L89
Haiman, Z. & Quataert, E. 2004, in Supermassive Black Holes in the Distant Universe. Edited by Amy J. Barger, Astrophysics and Space Science Library Volume 308. ISBN 1-4020-2470-3 (HB), ISBN 1-4020-2471-1 (e-book). Published by Kluwer Academic Publishers, Dordrecht, The Netherlands, p.147
Haller, J.W., Rieke, M.J., Rieke, G.H., Tamblyn, P., Close, L. & Melia, F. 1996, ApJ, 456, 194
Hamaus, N., Paumard, T., Müller, T., Gillessen, S., Eisenhauer, F., Trippe, S.& Genzel, R. 2009, ApJ, 692, 902
Harrison, C. M., Alexander, D. M., Swinbank, A. M., et al. 2012, MNRAS, 426, 1073
Harrison, C. M., Alexander, D. M., Mullaney, J. R., & Swinbank, A. M. 2014, MNRAS, 441, 3306
Heckman, T. M., Kauffmann, G., Brinchmann, J., Charlot, S., Tremonti, C. & White, S. D. M. 2004, ApJ, 613, 109
Heckman, T.M. 2010, in Co-Evolution of Central Black Holes and Galaxies, Proceedings of the International Astronomical Union, IAU Symposium, Volume 267, p. 3-14
Heckman, T.M. & Best, P.N. 2014, ARAA in press (astro-ph 1403.4620)
Herrnstein, J.T., Moran, J.M., Greenhill, L.J. & Trotter, A.S. 2005, ApJ, 629, 719
Hickox, R. C., Mullaney, J. R., Alexander, D. M., Chen, C.-T. J., Civano, F.M., Goulding, A. D., & Hainline, K. N. 2014, ApJ, 782, 9
Ho, L.C., Filippenko, A.V. & Sargent, W.L.W., 1997, ApJS, 112, 315
Ho, L.C. 2008, ARAA, 46, 475
Hopkins, P.F. et al. 2005, ApJ, 630, 716
Hopkins, P.F. 2006, ApJS, 163, 1
Hopkins, P.F. & Quataert, E. 2010, MNRAS, 407, 1529
Jahnke, K, et al. 2009, ApJ, 706, L215
Jahnke, K. & Maccio, A.V., 2011, ApJ, 743, 92
Kaspi, S., Smith, P. S., Netzer, H., Maoz, D., Jannuzi, B. T. & Giveon, U. 2000, ApJ, 533, 631
Kauffmann, G. & Haehnelt, M. 2000, MNRAS, 311, 576
Kauffmann, G. et al. 2003, MNRAS, 346, 1055
Kennea, J.A. et al.2013, ApJ, 770, L24
Kerr, R.1963, Ph.Rev.Lett., 11, 237
Kocevski, D.D. et al., 2012, ApJ, 744, 148
Korista, K. T., Bautista, M. A., Arav, N., et al. 2008, ApJ, 688, 108
Kormendy, J. 2004, in ‘Coevolution of Black Holes and Galaxies’, Carnegie Observatories Centennial Symposia. Cambridge University Press, Ed. L.C. Ho, p. 1
Kormendy, J., Bender, R. & Cornell, M.E. 2011 Nature 469, 374
Kormendy, J. & Ho, L. 2013, ARAA 51, 511
Krabbe, A. et al. 1995, ApJ, 447, L95
Kuo, C.Y., Braatz, J.A., Condon, J.J. et al. 2011, ApJ, 727, 20
Lacy, J.H., Townes, C.H., Geballe, T.R. & Hollenbach, D.J. 1980, ApJ 241, 132
Lawson, P.R., Unwin, S.C. & Beichman, C.A. 2004, JPL publication 04-014

Li, C., Kauffmann, G., Heckman, T. M., White, S. D. M.& Jing, Y. P. 2008, MNRAS, 385, 1915
Liebling, S.L. & Palenzuela, C. 2012, LRR, 15, 6
Liu, K., Wex, N., Kramer, M., Cordes, J. M.& Lazio, T. J. W. 2012, ApJ, 747,1
Lynden-Bell, D. 1969, Nature 223, 690
Lu, R.-S. et al. 2014, ApJ 788, L120
Macchetto, F., Marconi, A., Axon, D.J., Capetti, A., Sparks, W. & Crane, P. 1997, ApJ, 489, 579
Madau, P. et al. 1996, MNRAS, 283, 1388
Magorrian, J. et al. 1998, AJ, 115, 2285
Maldacena, J. 1998, Ad. Th.Math.Phys. 2, 231
Maoz, E. 1995, ApJ, 447, L91
Maoz, E., 1998, ApJ 494, L181
Marconi, A. et al. 2003, ApJ, 586, 868
Marconi, A., Risaliti, G., Gilli, R., Hunt, L. K., Maiolino, R.& Salvati, M. 2004, MNRAS, 351, 169
Marconi, A., Pastorini, G., Pacini, F., Axon, D. J., Capetti, A., Macchetto, D., Koekemoer, A. M. & Schreier, E. J.2006, A&A, 448, 921
Markoff, S., Falcke, H., Yuan, F. & Biermann, P.L. 2001, Astr.&Ap. 379, L13
Martini, P., 2004, in Coevolution of Black Holes and Galaxies, from the Carnegie Observatories Centennial Symposia, Cambridge University Press, as part of the Carnegie Observatories Astrophysics Series, edited by L. C. Ho, p. 169
Mayer, L., Kazantzidis, S., Escala, A.& Callegari, S. 2010, Nature, 466, 1082
McClintock, J. & R. Remillard 2004, in Compact Stellar X-ray sources , eds. W.Lewin and M.van der Klis, Cambirdge Univ. Press (astro-ph 0306123)
McConnell, N. & Ma, C.-P. 2013, ApJ 764, 184
McGinn, M.T., Sellgren, K., Becklin, E.E. & Hall, D.N.B. 1989, ApJ, 338, 824
McNamara, B.R. & Nulsen, P.E.J. 2007, ARAA, 45, 117
Merloni, A. et al. 2010, ApJ, 708, 137
Merritt, D., Alexander, T., Mikkola, S. & Will, C.M. 2010, PhRev D, 81, 2002
Meyer, L., Ghez, A. M., Schödel, R., Yelda, S., Boehle, A., Lu, J. R., Do, T., Morris, M. R., Becklin, E. E.& Matthews, K. 2012, Sci, 338, 84
Mori, K. et al. 2013, ApJ, 770, L23
Moscibrodzka, M., Falcke, H., Shiokawa, H. & Gammie, C. F. 2014, arXiv:1408.4743
Moster, B. P., Naab, T., & White, S. D. M. 2013, MNRAS 428, 3121
Miyoshi, M. et al. 1995, Nature 373, 127
Moran, J.M. 2008, ASPC, 395, 87
Mullaney, J.R , Daddi, E., Béthermin, M., Elbaz, D., Juneau, S., Pannella, M., Sargent, M. T., Alexander, D. M., Hickox, R. C. 2012, ApJ, 753, L30
Munyaneza, F, Tsiklauri, D. & Viollier, R.D. 1998, ApJ, 509, L105
Nandra, K., George, I. M., Mushotzky, R. F., Turner, T. J.& Yaqoob, T. 1997, ApJ, 477, 602
Nandra, K., O'Neill, P. M., George, I. M.& Reeves, J. N. 2007, MNRAS, 382, 194
Nesvadba, N. P. H., Polletta, M., Lehnert, M. D., et al. 2011, MNRAS, 415, 2359
Nesvadba, N. P. H., Lehnert, M. D., De Breuck, C., Gilbert, A. M. & van Breugel, W. 2008, A&A, 491, 407

Netzer, H. & Peterson, B.M. 1997, ASSL, 218, 85
Netzer, H., Mainieri, V., Rosati, P. & Trakhtenbrot, B. 2006, A&A, 453, 525
Netzer, H. 2009, MNRAS, 399, 1907
Netzer, H. 2013, The Physics and Evolution of Active Galactic Nuclei,Cambridge University Press
Neumayer, N., Cappellari, M., Rix, H.-W., Hartung, M., Prieto, M. A., Meisenheimer, K.& Lenzen, R. 2006, ApJ, 643, 226
Novak,G.S., Ostriker, J.P. & Ciotti, L. 2011, ApJ, 737, 26
Osmer. P.S. 2004, in Coevolution of Black Holes and Galaxies, from the Carnegie Observatories Centennial Symposia. Published by Cambridge University Press, as part of the Carnegie Observatories Astrophysics Series. Edited by L. C. Ho, p. 324.
Özel, F., Psaltis, D., Narayan, R. & McClintock, J. E. 2010, ApJ 725, 1918
Paumard, T., Perrin, G., Eckart, A., Genzel, R., Lena, P., Schoedel, R., Eisenhauer, F., Mueller, T. & Gillessen, S. 2005, Astr.Nachrichten 326, 568
Peng, Y. et al. 2010, ApJ, 721, 193
Peterson, B.M. 1993, PASP, 105, 247
Peterson, B.M. 2003, ASPC, 290, 43
Pfahl, E. & Loeb, A. 2004, ApJ, 615, 253
Psaltis,D. & Johanssen, T. 2011, JPhCS, 283, 2030
Rees, M.J. & Ostriker, J.P. 1977 MNRAS 179, 541
Rees, M. 1984, Ann.Rev Astr.Ap. 22, 471
Reid, M.J. & Brunthaler, A. 2004, ApJ 616, 872
Reid, M.J. Braatz, J. A., Condon, J. J., Lo, K. Y., Kuo, C. Y., Impellizzeri, C. M. V.& Henkel, C. 2013, ApJ, 767, 154
Reid, M.J. et al. 2014, ApJ, 783, 130
Reines, A. E.;,Sivakoff, G. R., Johnson, K. E.& Brogan, C. L.2011, Nature, 470, 66
Reines, A.E., Greene, J.E. & Geha, M. 2013, ApJ, 775, 116
Reynolds, C.S. 2013, CQGra., 30, 4004
Remillard, R.A. & McClintock, J.E. 2006, ARAA 44, 49
Rieke, G.H. & Rieke, M.J. 1988, ApJ, 330, L33
Rosario, D. et al. 2012, A&A, 545, 45
Rosario, D. et al. 2013, ApJ, 771, 63
Rupke, D. S. N. & Veilleux, S. 2013, ApJ, 768, 75
Sanders, D. B. et al. 1988, ApJ, 325, 74
Sanders, R.H. 1998, MNRAS, 294, 35
Schödel, R. et al. 2002, Nature 419, 694
Schödel, R. et al. 2003, ApJ 596, 1015
Schmidt, M. 1963, Nature 197, 1040
Schunck, F.E. & Mielke, E.W. 2003, CQGra, 20, R301
Schawinski, K.,Treister, E., Urry, C. M., Cardamone, C. N., Simmons, B., Yi, S. K. 2011, ApJ, 727, L31
Schwarzschild, K., 1916, Sitzungsber. Preuss. Akad.Wiss., 424
Scoville, N.Z. & Norman, C.A. 1988, ApJ, 332, 163
Sellgren, K., McGinn, M.T., Becklin, E.E. & Hall, D.N. 1990, ApJ 359, 112
Serabyn, E. & Lacy, J.H. 1985, ApJ 293, 445
Shakura, N.I. & Sunyaev, R.A. 1973, A&A 24, 337

Shankar, F., Weinberg, D.H. & Miralda-Escude, J. 2009, ApJ, 690,20
Shao, L. et al. 2010, A&A, 518, L26
Shen, Z.Q., Lo, K.Y., Liang, M.C., Ho, P.T.P. & Zhao, J.H. 2005, Nature 438, 62
Sijacki, D. Vogelsberger, M., Genel, S., Springel, V., Torrey, P., Snyder, G., Nelson, D. & Hernquist, L. 2014, arXiv1408.6842
Soltan, A. 1982, MNRAS, 200, 115
Somerville, R., Hopkins, P. F., Cox, T. J., Robertson, B. E. & Hernquist, L. 2008, MNRAS, 391, 481
Spitler, L.G. et al. 2014, ApJ, 780, L3
Sturm, E., González-Alfonso, E., Veilleux, S., et al. 2011, ApJ, 733, L16
Susskind, L. 1995, JMP 36, 6377
Tacconi, L.J., Neri, R., Genzel, R., et al. 2013, ApJ, 768. 74
Tanaka, Y. et al. 1995, Nature, 375, 659
Torres,D.F., Capoziello, S. & Lambiase, G. 2000, PhRv D, 62, 4012
Townes, C. H.; Lacy, J. H.; Geballe, T. R. & Hollenbach, D. J. 1982, Nature 301, 661
Trakhtenbrot, B. & Netzer, H. 2012, MNRAS, 427, 3081
Tremaine, S. et al. 2002, ApJ, 574, 740
Uttley, P. et al. 2014, A&Arev, 22, 72
van der Marel, R.P. & vn den Bosch, F.C. 1998, AJ, 116, 2220
Vestergaard, M. 2004, ApJ 601, 676
Veilleux,S., Kim, D.-C. & Sanders, D.B. 1999, ApJ, 522, 113
Veilleux, S., Cecil, G. & Bland-Hawthorn, J. 2005, ARAA, 43, 769
Veilleux, S. et al. 2009, ApJS, 182, 628
Veilleux, S., Meléndez, M., Sturm, E., et al. 2013, ApJ, 776, 27
Viollier, R.D, Trautmann, D. & Tupper 1993, PhLB, 306, 79
Volonteri, M. 2010, A&AR, 18, 279
Volonteri, M. 2012, Sci, 337, 544
Weinberg, N. N., Milosavljevic, M. & Ghez, A. M. 2005, ApJ 622, 878
Westmoquette, M. S., Clements, D. L., Bendo, G. J., & Khan, S. A. 2012, MNRAS, 424, 416
Wharton, R. S., Chatterjee, S., Cordes, J. M., Deneva, J. S.& Lazio, T. J. W. 2012, ApJ, 753, 108
Wheeler, J.A. 1968, Amer.Scient. 56, 1
Wild, V., Heckman, T. & Charlot, S. 2010, MNRAS, 405, 933
Will, C.M. 2008, ApJ, 674, L25
Wollman, E. R.; Geballe, T. R.; Lacy, J. H.; Townes, C. H.; Rank, D. M.1977, ApJ 218, L103
Yu, Q & Tremaine, S. 2002, MNRAS, 335, 965
Zakamska, N.L. & Greene, J.E. 2013, MNRAS, 442, 784
Zucker, S., Alexander, T., Gillessen, S., Eisenhauer, F. & Genzel, R. 2006, ApJ 639, L21